\documentclass[twocolumn,english,aps,prb]{revtex4}
\usepackage[T1]{fontenc}
\usepackage[latin9]{inputenc}
\usepackage{units}
\usepackage{amsmath}
\usepackage{graphicx}

\providecommand{\tabularnewline}{\\}
\newcommand{\lyxdot}{.}

\usepackage{babel}

\begin{document}

\title{Influence of Magnetism on Phonons in CaFe$_{\text{2}}$As$_{\text{2}}$ Via Inelastic X-ray Scattering}

\author{S. E. Hahn}

\email{shahn@ameslab.gov}

\affiliation{Ames laboratory and Department of Physics and Astronomy, Iowa State
University, Ames, IA, 50010}

\author{A. Alatas}

\affiliation{Advanced Photon Source, Argonne National Laboratory, Argonne, IL
60439}

\author{B. M. Leu}

\affiliation{Advanced Photon Source, Argonne National Laboratory, Argonne, IL
60439}

\author{Y. Lee}

\affiliation{Ames laboratory and Department of Physics and Astronomy, Iowa State
University, Ames, IA, 50010}

\author{N. Ni}

\affiliation{Ames laboratory and Department of Physics and Astronomy, Iowa State
University, Ames, IA, 50010}

\author{D. Y. Chung}

\affiliation{Materials Science Division, Argonne National Laboratory, Argonne,
IL 60439}

\author{I. S. Todorov}

\affiliation{Materials Science Division, Argonne National Laboratory, Argonne,
IL 60439}

\author{M. G. Kanatzidis}

\affiliation{Materials Science Division, Argonne National Laboratory, Argonne,
IL 60439}

\affiliation{Department of Chemistry, Northwestern University, Evanston, IL 60208}

\author{E. E. Alp}

\affiliation{Advanced Photon Source, Argonne National Laboratory, Argonne, IL
60439}

\author{P. C. Canfield}

\affiliation{Ames laboratory and Department of Physics and Astronomy, Iowa State
University, Ames, IA, 50010}

\author{A. I. Goldman}

\affiliation{Ames laboratory and Department of Physics and Astronomy, Iowa State
University, Ames, IA, 50010}

\author{R. J. McQueeney}

\email{mcqueeney@ameslab.gov}

\affiliation{Ames laboratory and Department of Physics and Astronomy, Iowa State
University, Ames, IA, 50010}

\author{B. N. Harmon}

\affiliation{Ames laboratory and Department of Physics and Astronomy, Iowa State
University, Ames, IA, 50010}
\begin{abstract}
In the iron pnictides, the strong sensitivity of the iron magnetic
moment to the arsenic position suggests a significant relationship
between phonons and magnetism. We measured the phonon dispersion of
several branches in the high temperature tetragonal phase of $\mathrm{CaFe_{2}As_{2}}$
using inelastic x-ray scattering on single-crystal samples. These
measurements were compared to $ab\ initio$ calculations of the phonons.
Spin polarized calculations imposing the antiferromagnetic order present
in the low temperature orthorhombic phase dramatically improve agreement
between theory and experiment. This is discussed in terms of the strong
antiferromagnetic correlations that are known to persist in the tetragonal
phase. 
\end{abstract}
\maketitle
Iron-pnictide materials are currently the subject of enormous scientific
activity. The discovery of superconductivity with $T_{c}\mathrm{'s}$
up to 55K and the ensuing efforts to understand the possible electronic,
magnetic and phononic mechanisms responsible have already revealed
intriguing information.\cite{Kamihara-3296,0256-307X-25-6-080} For
example, first principles calculations using density functional theory
for LaFeAsO strongly suggest that electron-phonon coupling is not
sufficient to explain superconductivity.\cite{boeri:026403} Other
calculations demonstrate strong dependence of the iron magnetic moment
on the arsenic atomic position suggesting an interplay between certain
phonons and magnetism.\cite{Mazin:nat_phys,yildirm-2009,mittal-2009} Experiments
and calculations of the crystal structure and magnetism under apressure
for $\mathrm{CaFe_{2}As_{2}}$ show a dramatic sensitivity of the
magnetic moment to changes in the atomic positions.\cite{kreyssig:184517}
The motivation for the present investigation was to study phonons
in $\mathrm{CaFe_{2}As_{2}}$ to ascertain any anomalies related to
the structure and possible magnetic interactions. 

We report measurements of the phonon dispersion in the paramagnetic
high-temperature tetragonal phase of single-crystal CaFe$_{2}$As$_{2}$
using inelastic x-ray scattering. Several phonon branches consisting
of c-axis polarized As modes are observed to have energies and intensities
in poor agreement with non-spin-polarized (NSP) band structure calculations
of the phonon dispersion and structure factor. However, the imposition
of antiferromagnetic ordering in the tetragonal phase by spin-polarized
(SP) calculations brings the dispersion into better agreement with
the experimental data. Strong antiferromagnetic correlations have been 
observed by neutron scattering above the AFM ordering temperature 
\cite{mcqueeney-2008} and extend above room temperature.\cite{mcqueeney-unpublished}
In addition, X. F. Wang et. al.\cite{wang:117005} measured a substantial 
linear increase in the magnetic susceptibility of $\mathrm{BaFe_{2}As_{2}}$
up to 700K.  Such behavior may indicate spin fluctuations above the AFM 
ordering temperature.\cite{Zhang-2009}  Similar behavior has also been reported for 
 $\mathrm{CaFe_{2}As_{2}}$ up to at least 300K.\cite{ni:214515}
Resistivity measurements also suggest spin fluctuations and spin scattering 
above the AFM transition temperature.\cite{tanatar:134528} Given the known 
sensitivity of the Fe magnetic moment to the As $z$-position in the crystal, 
the measurements and calculations of the phonon structure factor indicate 
an important coupling between local magnetic order and inter-atomic force 
constants.

The $\mathrm{CaFe_{2}As_{2}}$ sample used for measurements along
(110) was prepared by a stoichiometric elemental mixture in Sn metal
flux. Handling of all elements was carried out in a $\mathrm{N_{2}}$-filled
glove box and 0.2 grams of $\mathrm{Ca}$, Fe, and As and 4.0 grams
of Sn metal were used. The mixture was loaded into an alumina tube
plugged with ceramic wool and sealed in a silica tube under vacuum.
The mixture was heated to 1050$\mathrm{^{\text{o}}C}$ for 14 hours
and kept there for 8 hours, followed by cooling to 600$\mathrm{\mathrm{^{\text{o}}C}}$
at a rate of 20 $\mathrm{\unitfrac{^{o}C}{hour}}$ . The molten Sn
flux was then filtered with a centrifuge. The resulting square plate
crystals of $\mathrm{CaFe_{2}As_{2}}$ were identified by X-ray powder
diffraction. All other measurements were performed on samples that
were prepared as described elsewhere.\cite{ni:014523,canfield-2009}

Inelastic x-ray scattering measurements were performed at Sector-3-ID-C
of the Advanced Photon Source at Argonne National Laboratory. The
incident energy was set to 21.66 keV with an energy resolution of
2.379 meV full-width-at-half-maximum. Measurements were performed
on a single-crystal of CaFe$_{2}$As$_{2}$ measuring approximately
2 x 2 mm with a thickness of ~100$\mu$m. The sample was mounted
in both the $(H0L)$ and $(HHL)$ tetragonal planes in a closed-cycle
Helium refrigerator and most measurements were performed at room temperature,
although some scans were also performed at lower temperatures.

\begin{figure}
\begin{tabular}{c}
\includegraphics{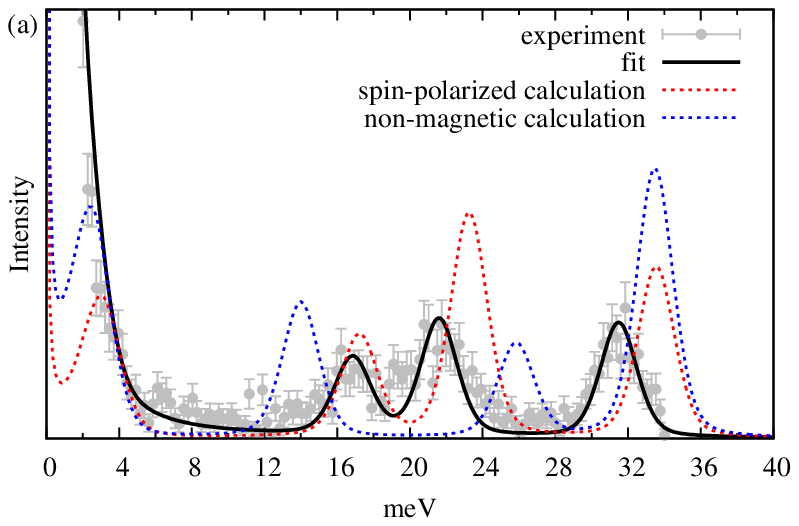}\tabularnewline
\includegraphics{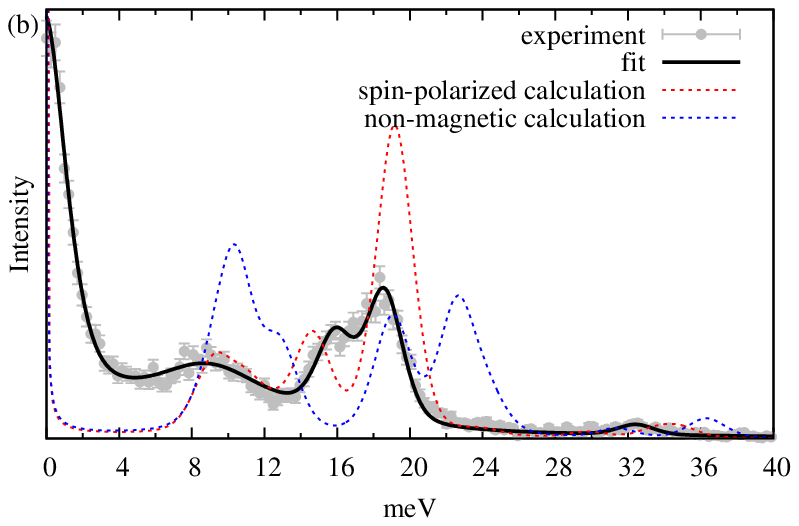}\tabularnewline
\end{tabular}\caption{(a) Constant-Q line scan at $Q=\left(0,0,5.75\right)$ . (b) Constant-Q
line scan at $Q=\left(1.5,0,3.0\right)$ . Experimental data are
given by points and pseudo-Voigt fits by the solid line. Dashed red
and blue lines correspond to calculations of the phonon structure
factor for spin-polarized and non-magnetic calculations, respectively.\label{fig:line scans}}

\end{figure}

Phonons were measured using constant-$Q$ energy scans along the tetragonal
$(0,0,5+\xi)$, $(1,0,3+\xi)$, $(1+\xi,0,3)$, and $(\xi,\xi,4)$ directions
at room temperature. The energy transfer scale for each scan was calibrated
for monochromator and analyzer temperature drifts. The scans were
fit to several peaks using a pseudo-Voigt line profile. The normalized
pseudo-Voigt function is given in eq.\ref{eq: pseudo-Voigt}, where
$f_{G}\left(x;\Gamma\right)$and $f_{L}\left(x;\Gamma\right)$ are
normalized Gaussian and Lorentzian functions respectively. The mixing
parameter $\eta=0.393$, and resolution full-width-at-half-maximum
(FWHM) $\Gamma=2.379\ meV$ was determined from fits to the elastic
scattering width of Plexiglas. \begin{equation}
f_{pV}=\left(1-\eta\right)f_{G}\left(x;\Gamma\right)+\eta f_{L}\left(x;\Gamma\right)\label{eq: pseudo-Voigt}\end{equation}
 Figures \ref{fig:line scans}(a) and \ref{fig:line scans}(b) show
line scans consisting of several phonon excitations at $Q=\left(0,0,5.75\right)$
and $Q=\left(1.5,0,3\right)$ respectively along with fits to the
data. The peak positions for these and other scans were obtained from
the fits and used to construct the dispersion of phonon branches along
the different scan directions.

In order to understand the features of the phonon dispersion, the
experimental measurements were compared to $\mathit{ab\ }initio$
calculations of the phonons. The phonon dispersion was calculated
using Density Functional Perturbation Theory (DFPT).\cite{PWSCF}
There is not yet a consensus on the proper lattice parameters to use for these calculations. 
In nonmagnetic calculations, relaxing the lattice parameters results in a large contraction
of the c-axis. In spin-polarized calculations, the lattice distorts into an orthorhombic 
structure. With these difficulties in mind, experimental lattice parameters were used.\cite{kreyssig:184517} 
In addition, there is controversy over the appropriate internal Arsenic parameter to use. 
We chose relaxed positions so that all forces were zero. 
For non-magnetic and spin-polarized calculations, the relaxed As z-position
is $z_{As}=0.3575$ and $z_{As}=0.3679$ respectively. Mittal et. al.\cite{mittal-2009} use
the experimental value of $z_{As}=0.372(1)$. The pseudopotentials
chosen used the Perdew-Burke-Ernzerhof (PBE) exchange correlation
functional.\cite{PSEUDO,PhysRevLett.77.3865} Settings of an 8x8x8
k-mesh and 24 eV and 400 eV energy cutoffs for the wavefunctions and
charge density were chosen to ensure that the precision of the calculated
phonon dispersion was better than $1.0\,\mathrm{meV}$. Due to the
paramagnetic state of CaFe$_{2}$As$_{2}$ at room temperature, non-magnetic
calculations were first performed. Phonon frequencies were calculated
on a 2x2x2 q-mesh, and then interpolated along several symmetry directions.
The resulting phonon frequencies and eigenvectors were used to calculate
the dynamical structure factor along the selected scan directions.
The dynamical structure factor, which is proportional to the x-ray
scattering intensity, is given in eq.\ref{eq: structure factor 1}.\cite{lovesey1984}
The Debye-Waller, $W_{d}\left(\mathbf{Q}\right)$ factor was set equal
to zero, and the scattering length is proportional to $Z_{d}$, the
atomic number of the corresponding atom, and $\mathbf{\sigma}_{\mathbf{d}}^{j}(\mathbf{q})$
is the eigenvector corresponding to the normalized motion of atom
d in the $\mathrm{j^{th}}$ phonon branch.

\begin{eqnarray}
S_{j}\left(\mathbf{q},\omega\right)=\frac{\left|H_{\mathbf{q}}^{j}\left(\mathbf{Q}\right)\right|^{2}}{2\omega_{j}\left(\mathbf{q}\right)}\left(1+n_{j}\left(\mathbf{q}\right)\right)\delta\left\{ \omega-\omega\left(\mathbf{q}\right)\right\} \label{eq: structure factor 1}\\
H_{\mathbf{q}}^{j}\left(\mathbf{Q}\right)=\sum_{d}\frac{Z_{d}}{\sqrt{M}_{d}}\exp\left(-W_{d}\left(\mathbf{Q}\right)+i\mathbf{Q}\cdot\mathbf{d}\right)\left\{ \mathbf{Q\cdot\mathbf{\sigma_{d}^{j}\left(\mathbf{q}\right)}}\right\} \label{eq: structure factor 2}\end{eqnarray}

\begin{figure}
\includegraphics{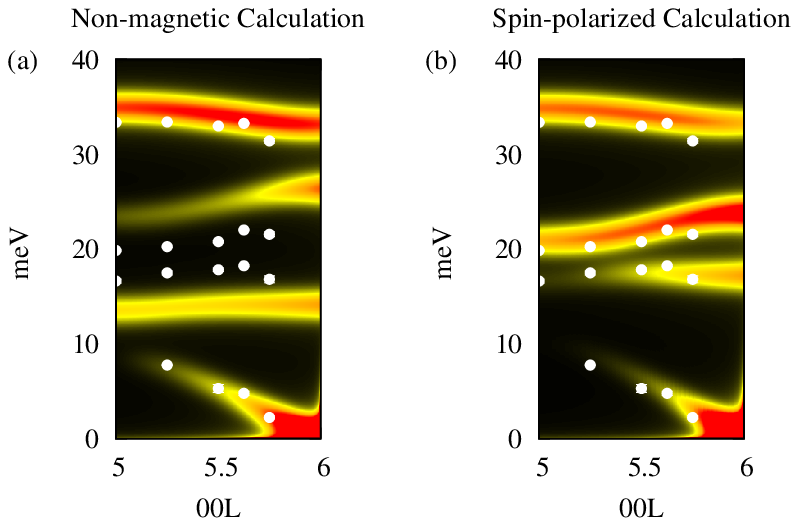}

\caption{Phonon dispersion weighted by the structure factors along (00L) (a)
without and (b) with antiferromagnetic order. The white dots are experimental
data points.\label{fig: 00L dispersion}}

\end{figure}

\begin{figure}
\includegraphics{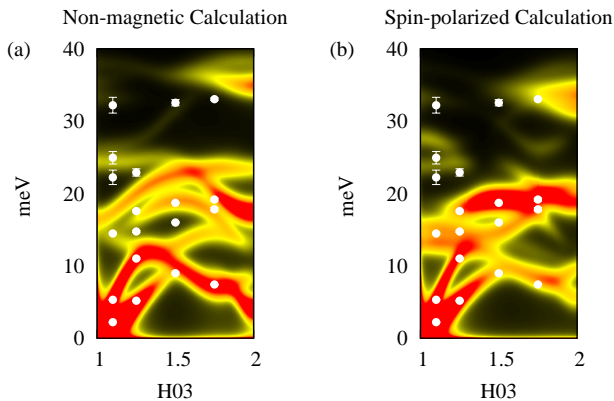}\caption{Phonon dispersion weighted by the structure factors along (H03) (a)
without and (b) with antiferromagnetic order. The white dots are experimental
data points.\label{fig: H03 dispersion}}

\end{figure}

For comparison with experiment, the delta functions were broadened
with a pseudo-Voigt function whose parameters were chosen to match
the experimental resolution. Fig.\ref{fig:line scans} shows the comparison
of the calculated structure factor to the data for two scans. For
the scans shown, the non-magnetic calculations show poor agreement
with the phonons observed near $\sim$ 17 and 20 meV. Figs.\ref{fig: 00L dispersion}(a)
and \ref{fig: H03 dispersion}(a) show several observed phonons (white
dots) along $(00L)$ and $\left(H03\right)$ compared to the non-magnetic
calculated phonon dispersion weighted by the structure factors.

The agreement between calculations and the data dramatically improve
when spin-polarized calculations were performed using a 4x4x4 k-mesh
in a supercell using the observed {}``stripe'' AF structure of orthorhombic
CaFe$_{2}$As$_{2}$. PBE overestimates the magnetic moment, with
a value of 2.39 $\mu_{B}/\mathrm{Fe}$ versus 0.80(5) $\mu_{B}/\mathrm{Fe}$
observed experimentally.\cite{goldman-2008-78} Similar discrepancies between the 
calculated and observed magnetic moment were discussed in phonon studies of 
$\mathrm{BaFe_{2}As_{2}}$.\cite{zbiri:064511} Software limitations prevented constraining 
the magnetic moments while performing phonon calculations. Fig.\ref{fig:line scans} shows that 
the spin-polarized calculations not only provide better agreement with the phonon energies, but 
even the phonon intensities (eigenvectors) show marked improvement. Along $\left(00L\right)$
in fig.\ref{fig: 00L dispersion}(b), the agreement of the middle
two branches at 17 and 20 meV is dramatically improved by including
the AF ordering which reduces the energy splitting of these two branches.
Examination of the eigenvectors of these modes shows that they have
$\Lambda_{1}$ symmetry and consist of longitudinally polarized Ca
and As modes. SP calculations also show marked improvement for phonon
branches in the same energy range along the (100) direction which
connect to the two $\Lambda_{1}$ branches at the zone boundary. Fig.\ref{fig:line scans}(b)
shows a scan at (1.5,0,3) where SP calculations dramatically improve
agreement with the two $\Delta_{3}$ phonon branches consisting of
c-axis polarized transverse Ca and As modes in the 14-20 meV range.
Fig.\ref{fig: H03 dispersion}(b) also shows improved agreement of
the SP calculations of select phonon branches along (H03) to the measured
dispersion. Measurements of other phonon branches along (110) and
(10L) also show better agreement with the SP calculations (not shown).

Along both $\left(00L\right)$ and $\left(H00\right)$, the upper
branch that softens with the introduction of AF order is associated
with c-axis polarized As vibrations. Such modes have been predicted
to have a strong influence on magnetism due to the sensitivity of
the Fe moment on the As $z$-position.\cite{yildirm-2009} Surprisingly,
the lower branch starting at 17 meV are primarily c-axis polarized
Ca modes in the NSP calculations. However, the spin-polarization introduces
a strong mixing of these two branches of identical symmetry. Such
mixing leads to the increased frequency and changes in intensity between
SP and NSP calculations shown in figs.\ref{fig: 00L dispersion} and
\ref{fig: H03 dispersion}. 

The improved agreement of the spin-polarized calculations in the paramagnetic
phase may seem surprising, however resistivity and magnetic susceptibility measurements
suggest spin fluctuations and spin scattering above the AFM transition temperature.
Inelastic neutron scattering experiments indicate substantial antiferromagnetic correlations 
that persist above $T_{N}=172K$\cite{mcqueeney-2008} and up to at least 300 K.\cite{mcqueeney-unpublished}
These correlations are observed up to high energies ( > 60 meV) with
correlation lengths up to 20\AA, which indicates that strong AF correlations
exist between sizable Fe moments even above $T_{N}$. Magnetism
must be accounted for when considering the chemical binding and the
interatomic forces. Cooling of the sample into the AF ordered orthorhombic
phase at 140 K appears to have little influence on the position or
linewidth of the phonons along $\left(00L\right)$.

In summary, we have measured the phonon dispersion along several high
symmetry directions for the paramagnetic high temperature tetragonal
phase of $\mathrm{CaFe_{2}As_{2}}$. Spin-polarized first principles
calculations are in better agreement with the experimental results
than non-spin-polarized calculations. The effects of large theoretical
magnetic moments and the details of the spin fluctuations at room
temperature are issues still under investigation. 

At the end of these investigations, Mittal, et. al.\cite{mittal-2009}posted
data from inelastic neutron scattering experiments and non-spin-polarized
band structure calculations for $\mathrm{CaFe_{2}As_{2}}$. This measurement 
is in excellent agreement with their published dispersion curves.
Considerations of magnetic effects on the phonon force constants and 
density of states were posted by I. I. Mazin and M. D. 
Johannes\cite{Mazin:nat_phys} and T. Yildirim.\cite{yildirm-2009}

\begin{acknowledgments}
Work at the Ames Laboratory was supported by the U.S. Department of
Energy, Basic Energy Sciences under Contract No. DE-AC02-07CH11358.
Use of the Advanced Photon Source was supported by the U. S. Department
of Energy, Office of Science, Office of Basic Energy Sciences, under
Contract No. DE-AC02-06CH11357 and LDRD program at Argonne National
Laboratory. 

\end{acknowledgments}

\bibliographystyle{apsrev}
\bibliography{CaFe2As2_phonons}

\end{document}